\documentclass{PoS}
\title{Coulomb gauge Green functions and Gribov copies in $SU(2)$
      lattice gauge theory\thanks{Supported by DFG under contract
      \textsl{DFG-Re856/6-1,2}.}}

\ShortTitle{Coulomb gauge Green functions and Gribov copies in $SU(2)$
      lattice gauge theory}

\author{\speaker{Markus Quandt}\\
        University of T\"ubingen\\
        E-mail: \email{quandt@tphys.physik.uni-tuebingen.de}}

\author{Giuseppe Burgio\\
        University of T\"ubingen\\
        E-mail: \email{burgio@tphys.physik.uni-tuebingen.de}}

\author{Songvudhi Chimchinda\\
        University of T\"ubingen\\
        E-mail: \email{chimchinda@tphys.physik.uni-tuebingen.de}}

\author{Hugo Reinhardt\\
        University of T\"ubingen\\
        E-mail: \email{hugo.reinhardt@uni-tuebingen.de}}

\abstract{We reconsider the lattice measurement of Green functions in
Coulomb gauge, both in $2+1$ and $3+1$ dimensions, using an improved gauge
fixing scheme. The influence of Gribov copies is examined and we find clear
indications that Green functions are more strongly affected than previously
assumed, in particular for low momenta. Qualitatively, our improved lattice
results in the infra-red compare more favourably with recent continuum
calculations in the Hamiltonian approach.}

\FullConference{The XXV International Symposium on Lattice Field Theory\\
                 July 30 - August 4 2007\\
                 Regensburg, Germany}

\newcommand{\tr}{\mbox{tr}}

\newcommand{\vek}[1]{\mathbf{#1}}

\usepackage{graphicx}

\begin{document}

%%%%%%%%%%%%%%%%%%%%%%%%%%%%%%%%%%%%%%%%%%%%%%%%%%%%%%%%%%%%%%%%%%%%%%%%%%%%%%%%

\section{Introduction}

Yang-Mills theory in the Coulomb gauge has recently drawn a renewed
attention, both in the continuum \cite{claus_hugo, szepaniak} and on the
lattice \cite{kurt,cucc,voigt}. In the continuum at least, this interest
is mostly due to the remarkable fact that Gau{\ss}' law can be resolved
explicitly in Coulomb gauge, which gives the remaining vector potential
$\vek{A}$ a very intuitive notion similar to electrodynamics \cite{zwanziger}.
Recent variational approaches in the Schr\"odinger picture even support
the idea of a \emph{constituent gluon} \cite{claus_hugo, szepaniak},
which is almost non-interacting in the infrared and thus completely
determined by its dispersion relation $\omega(\vek{p})$, i.e.~the (inverse)
equal-time gluon propagator $D(\vek{p}) = \frac{1}{2}\omega(\vek{p})^{-1}$.

The obvious drawback of the Coulomb gauge is that \emph{manifest} Lorentz
invariance is lost at intermediate stages; it may only be recovered at
the end of the calculation. Perturbatively, this problem is reflected in the
(tree-level) propagators of some fundamental fields,  which are instantaneous
in time so that many loop integrands are independent of the temporal
loop momentum component $k_0$. Such integrals are notoriously difficult
to regulate with conventional techniques, though they are believed to cancel
in the full theory \cite{zwanziger}. Still, the issue of renormalisation
in Coulomb gauge remains cumbersome, even at the one-loop level
\cite{peter}.

Similar problems arise on the lattice as well. While initial studies
of the gluon propagator in Coulomb gauge displayed almost perfect
scaling \cite{kurt,cucc}, recent studies using improved gf.~techniques
indicate that the quality of gauge fixing has a significant impact
on Green functions; in particular, substantial scaling violations may
result \cite{voigt}. The same conclusion has been drawn earlier in
Landau gauge, where careful gauge fixing may alter the infrared
behaviour of the propagator quantitatively by as much as 20 \%
\cite{mueller_preusker}.

Even more severe descrepancies arise in the comparision of
early lattice results with the variational approach mentioned above.
While both methods show good agreement in $D=2+1$, their results
in $D=3+1$ differ \emph{qualitatively}, both in the infra-red and the
ulta-violet:

\begin{center}
\begin{tabular}{|c||c|c|}\hline
 & IR & UV \\\hline\hline
lattice  \cite{kurt,cucc} \rule[-3mm]{0mm}{8mm}& $D(\vek{p}) \to \mbox{const}$ &
$D(\vek{p}) \sim |\vek{p}|^{-\frac{3}{2}}$ \\\hline
variation \cite{claus_hugo}\rule[-3mm]{0mm}{8mm}&
$D(\vek{p}) \to 0$ & $D(\vek{p}) \sim |\vek{p}|^{-1}$ \\\hline
\end{tabular}
\end{center}

All these findings emphasise the need for a thorough corroboration
of lattice results in Coulomb gauge, in particular with regard to the
quality of gauge fixing. In the present talk, I will present the first
results in this program, viz.~the equal time gluon propagator in
$D=2+1$ and $D=3+1$. Further studies on the ghost propagator and
the Coulomb form factor are currently underway
and will be presented elsewhere.

The plan of this talk is as follows: In the next section, I will briefly
discuss our gf.~techniques and  demonstrate that they are effective in
reducing the Gribov problem which is at the heart of most gf.~issues.
Section three presents our findings for the gluon propagator. Some
of this data is still preliminary, and so is the quantitative
analysis, but our results so far imply both scaling violations
in the UV and a significant suppression in the IR. The last
point improves the qualitative agreement with variational studies,
although the quantiative agreement is still unsatisfactory.
In the last section, I will conclude with a brief summary and
outlook.

\section{Gauge fixing techniques}

\noindent
Coulomb gauge on the lattice can be defined as the maximisation
of the functional\footnote{For simplicity, we work exclusively
with the colour group $G=SU(2)$.}
\begin{equation}
F_t[U] \equiv \frac{1}{3\,V_3}\,\mathrm{tr}\,\sum_{\vek{x}}\sum_{i=1}^3
\,\frac{1}{2} \,\tr\, U_i(\vek{x},t) \stackrel{!}{=} \max\,,\qquad\quad
V_3 \equiv \prod_{i=1}^3 N_i \,.
\label{1}
\end{equation}
Here, $U_\mu(x)$ are the link variables, the sum over $\vek{x}$ runs
over all sites in a fixed  \emph{time-slice} $t = \mbox{const}$
and the maximisation is along the gauge orbit, i.e.~with respect to
all gauge rotations $\Omega(\vek{x},t)$ of the link field
$U_\mu(x)$.
As indicated, the Coulomb condition $F_t \stackrel{!}{=} \max$
can be implemented at each time-slice $t$ \emph{independently}. This
leaves a residual invariance of space-independent but time dependent
gauge transformations $\Omega(t)$, i.e.~a global gauge rotation in every
time slice.

For the equal-time gluon propagator\footnote{Gauge potentials are
extracted from the link variables in the usual fashion via an
$\mathcal{O}(a^2)$ improvement of the basic formula
$A_\mu = \frac{1}{2 a} \left[ U_\mu(x) - U_\mu^\dagger(x)\right]$.}
\begin{equation}
D(\vek{p}) \sim \int d^3 \vek{x}\, e^{i \,\vek{p}\cdot(\vek{x}-\vek{y})}
\sum_{i=1}^3 \sum_{c=1}^3
\,\langle A_i^c(\vek{x},t)\,A_i^c(\vek{y},t) \rangle =
|\vek{p}|^{-1} + \mathcal{O}(\hbar)
\end{equation}
the residual gauge fixing is irrelevant and it is sufficient to
fix only the time slice in which the measurement is taken. This is
no longer true for other correlators such as the $A_0 - A_0$ propagator
related to the static Coulomb potential. Moreover, recent perturbative
studies \cite{peter} indicate that possible scaling violations in
$D(\vek{p})$ may be attributed to the loss of covariance at equal times;
it will then be necessary to consider the full gluon propagator at all
(unequal) times, and Coulomb gauge fixing at all time slices must be
augmented by a suitable choice for the residual symmetry.

% Although not directly of relevance for the studies reported in this talk,
% the issue of the residual symmetry left after Coulomb gauge fixing
% can be discussed in terms of the adjoint field
% \begin{equation}
% \phi^c(t) \equiv \frac{1}{V_3}\,\int d^3\vek{x}\, A_0^c(\vek{x},t)\,.
% \label{3}
% \end {equation}
% In the continuum (and in particular in the variational approaches which
% choose the Weyl gauge $A_0=0$ from the outset), we want
% $\phi = 0$. However, this requirement cannot be directly translated to
% the lattice: Apart from the issue of inequivalent discretisations
% of (\ref{3}) the periodic boundary conditions prevent $\phi=0$.
% At most, one can set $\phi(t) = \mbox{const}$
% and use global transformations to rotate $\phi$ in the Cartan-algebra.
% A more pragmatic approach is to let the residual gauge symmetry
% \emph{unfixed}. This leads to  $\langle \phi(t) \rangle = 0$, which may
% be a viable approximation to the continuum condition $\phi(t)=0$.
% In the present study, we adopt this prescription.
% Let me stress again that any other choice would not affect
% $D(\vek{p})$, although it will have an impact on other correlators,
% in particular on the full gluon propagator.

The Gribov problem, which is at the heart of most g.f.~issues, can be
expressed as the fact that (\ref{1}) has many \emph{local} maxima
which may, however, give inequivalent contributions to non-gauge invariant
quantities such as the Green functions.
% In the continuum, uniqueness can be enforced by restricting the gauge
% potential $A_\mu$ to the so-called fundamental modular region.
Uniqueness can be enforced by searching for the \emph{global} maximum of
(\ref{1}), an NP-hard problem. Our strategy to reduce the influence
of Gribov copies is to prepend the standard (over)relaxation algorithm
by an initial \emph{preconditioning} step combined with multiple
\emph{Gribov repetitions}  from random starts.
This method is a less expensive substitute for full simulated annealing
and works well for small to medium size volumina up to $V \approx36^4$.
% For even larger lattices,
% the global search capabilities of a simulated annealing type of
% method becomes undispensable, cf.~ref.~\cite{voigt}.

\subsection{Preconditioning}

The periodic boundary conditions on the lattice allow for a somewhat
larger symmetry than just the periodic local gauge rotations. This
is well-known from the $SU(2)$ lattice center symmtry: In this case, one
multiplies all links $U_0(t,\vek{x})$ pointing out of a fixed
time-slice\footnote{The actual location of the time slice $t$ is
irrelevant, since a center flip at a different time-slice $t'$
can be decomposed into a flip at $t$ followed by a strictly local,
periodic gauge transformation.} $t=\mbox{const}$ by $(-1)$. This
construction flips the sign of all Polyakov lines, but it leaves all
plaquettes (and thus the action)  invariant; it is therefore
a genuine symmetry of the system. In Landau gauge, one can generalise
this construction to all four directions, giving a total of $2^4$
possible combinations of \emph{flips} \cite{mueller_preusker}.

In the Coulomb case, the gauge fixing is carried out in a fixed
$3D$ time-slice, i.e.~the flips are only carried out in spatial
directions, and only the $2D$ sub-planes perpendicular to a
given direction (at fixed $t$) are flipped.
The \emph{preconditioning} consist of trying all $2^3$ twists to
maximise $F_t[U]$ prior to the acutal relaxation step.
This can be viewed as a non-local update representing a large
symmetry transformation  that no local relaxation algorithm is likely
to find. Flips can also be interspersed at any time during relaxation,
although they are most efficient early on, when the algorithm has not
yet converged onto a target maximum.\footnote{The (over)relaxation
algorithm is iterated until the local gf.~violation, i.e.~the (maximal
norm at all sites $\vek{x}$ of the) local gradient of (\ref{1}) is
smaller than $10^{-13}$.}

\subsection{Multiple Gribov repetitions}
\label{gr}
The gf.~sequence consisting of preconditioning, relaxation and
overrelaxation can be repeated multiple times with random
starting points. This inspects different regions of the search
space and converges to distinct Gribov copies.
What makes this repetition effective is
that a relatively small number $N$ of copies gives a large
increase in the gf.~functional, while subsequent repetitions
beyond a certain \emph{plateau} point do not give any substantial
improvement within reasonable computation time.

This can be seen in figure 1, which plots the equal-time
Gluon propagator $D(\vek{p})$ at the smallest non-zero lattice
momentum, as a function of the number $N$ of Gribov repetitions.
The net effect of the improved gauge fixing is generally to
suppress $D(\vek{p}_{\rm min})$. Even for $N$ as small as
$N=2,\ldots,5$, the corrections are in the range of $10 \%$.
Further copies give smaller corrections; it is then a matter of experiment
to find the optimal tradeoff between CPU time and gf.~quality.
The optimal $N$ will depend quite sensitively on the lattice size and
other simulation parameters. In fig.~1  one can see the plateau
setting in rather quickly, while our largest lattices
($V=36^4$) required up to $N=30$ repetitions.
%
% We have conducted extensive experiments to determine the optimal
% number of Gribov repetitions for all our  simulation parameters.

\begin{figure}
\vspace*{3mm}
\begin{center}
\includegraphics[width=9cm, height=5cm]{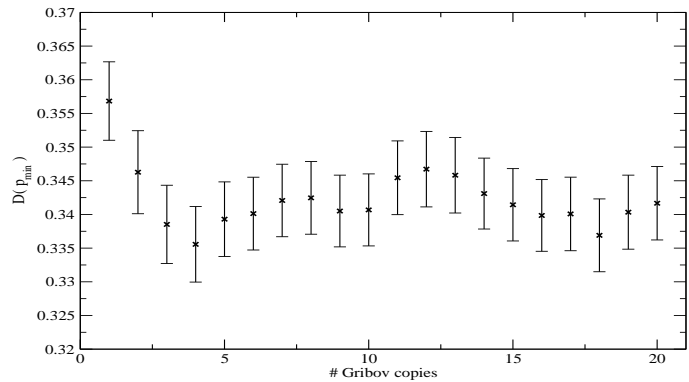}
%\hspace{.5cm}
%\includegraphics[width=7cm]{gcopf24_22.eps}
\end{center}
\label{fig1}
\caption{The equal-time gluon propagator at the smallest non-zero
lattice momentum, measured as a function of the number $N$ of
Gribov repetitions. Data was collected on a $24^4$ lattice with
$\beta=2.15$ (left) and $\beta=2.20$ (right); a total of $200$ thermalised
configuration were analysed for each data point.}
\end{figure}

%%%%%%%%%%%%%%%%%%%%%%%%%%%%%%%%%%%%%%%%%%%%%%%%%%%%%%%%%%%%%%%%%%%%%%%%%%%%%%%
\section{Results}

\subsection{$D=2+1$}

In this case, our findings in fig.~2 are in fair agreement to previous
lattice calculations \cite{kurt, cucc}. Our improved gauge fixing
scheme has again the tendency to suppress the gluon propagator
in the infra-red, but since $D(|\vek{p}|) \to 0$ at small $|\vek{p}|$ even
without gf.~improvment, the \emph{qualitative} behaviour of the
gluon propagator is unchanged.

In the UV, we observe scaling in the sense that the various propagator
curves for different values of the coupling $\beta$ can be multiplied by
a momentum-independent factor $Z(\beta)$ such that all curves coalesce to a
single line. There is a tendency for the scaling to be less perfect than
without the gf.~improvement, but this is well below the error bars of
our numerical simulation.

Quantitatively, the suppression of the gluon propagator in the infra-red
is as large as $10 \%$ -- $15 \%$. To fit the curve in the deep IR and
UV region, we have placed two cuts on the data.
In the IR, a power ansatz yields
\begin{equation}
D(|\vek{p}|) = |\vek{p}|^\alpha \cdot (c_1 + c_2 |\vek{p}|^2 + \cdots)\,,
\qquad\qquad \alpha \approx 0.85(10).
\end{equation}
Since the curve flattens towards the maximum, the exponent $\alpha$ is
somewhat depending on the exact location of the IR cut. At
$\Lambda = 0.5\,\mathrm{GeV}$, we have $\alpha = 0.81$, while it increases
to the above value $\alpha = 0.85$ for $\Lambda= 0.4 \,\mathrm{GeV}$.
With our present lattice sizes, we cannot go much lower with the IR
cut, but the present trend does certainly not rule out the value
$\alpha = 1$ preferred by Hamiltonian approaches \cite{claus_hugo}.

In the ultra-violet, a power-law decay
\begin{equation}
D(|\vek{p}|) \sim |\vek{p}|^{-\gamma}\,,\qquad \gamma \approx 1.5(1)
\end{equation}
is possible, but the exact value of the exponent $\gamma$ depends quite
sensitively on the location of the UV cut. A double-logarithmic plot
in the deep UV is \emph{not} a straight line at large momenta, which
points to sizeable logarithmic corrections. In fact, an ad-hoc ansatz
\[
D(p) \sim \frac{1}{ |\vek{p}| \cdot \ln |\vek{p}|^\delta}
\]
with $\delta \approx 0.51$ can fit the data equally well. The conclusion
is that our present data does not contain large enough momenta to distinguish
between a logarithmic or a power-like behaviour in the UV.

\begin{figure}
\begin{center}
\vspace*{4mm}
\includegraphics[width=9cm]{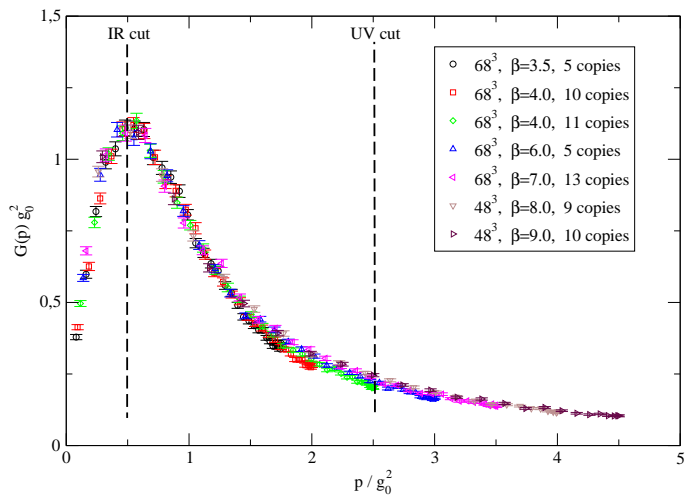}
\end{center}
\label{fig2}
\caption{The renormalised equal-time gluon propagator for various
couplings and lattice sizes. For the significance of the two data cuts,
see the main text.}
\end{figure}

\subsection{$D=3+1$}

The left panel of fig.~3 shows the results
for the largest lattice that we considered. The improved gf.~scheme
is now seen to make a \emph{qualitative} difference, both in the IR and
the UV.

At low momenta, the propagator is clearly \emph{suppressed} as compared
to less intricate gf.~procedures. The power-law fit explained in the last
subsection reveals a IR exponent of
\[
\alpha \approx 0.24(12)\,,
\]
again with significant variations as the IR cut on the data is changed.
However, a value $\alpha=0$, i.e.~a gluon propagator going to a constant
as $\vek{p} \to 0$ \cite{kurt, cucc} seems much more unlikely than
the vanishing $D(0) = 0$ predicted by variational calculations \cite{claus_hugo}.
On the other hand, the comparision with the $D=2+1$ case indicates that
much smaller momenta must be sampled to rule out one or the other option.
% In any event, such small momenta can only be studied on very large
% lattices, which in turn makes even more advanced simulated annealing type of
% gf.~techniques indispensable. Preliminary studies  \cite{voigt} in this
% direction agree with the suppression reported here, and the scenario
% $D(0) = 0$ at this point looks more likely.

In the UV, the most striking difference to previous lattice results is
the absence of perfect scaling, i.e.~the gluon propagator does not seem
to be multiplicatively renormalisable. This can be clearly seen in the
logarithmic plot in the right panel of fig.~3. In a multiplicatively
renormalisable situation, we would expect the curves for all
couplings $\beta$ to have the same \emph{slope} at large momenta --
which is clearly not the case.

One can now proceed and renormalise
anyway such that a common curve can be observed in one $p$-region or the
other (the right panel of fig.~3 has been renormalised to fit
well in the IR). In particular, one could try to fit the deep
UV region, at the expense of sacrificing a common curve in the IR.
From such a fit, it is even possible to extract a power-like behaviour
\[
D(\vek{p}) \sim |\vek{p}|^{-\alpha}\,,\qquad\qquad \alpha = 1.57\,.
\]
which is in fair agreement with ref.~\cite{kurt}. Our present data, however,
does not warrant such a procedure. In particular, an ad-hoc logarithmic
ansatz as  in the last subsection would work equally well.
To summarize, the scaling violations displayed by our improved gf.~scheme
are so severe that any attempt to extract a consistent UV behaviour
from a multiplicative renormalisation seems ill-adviced.

Comparable problems with renormalisation were also found in other studies
employing improved gf.~schemes. Continuum perturbation theory
\cite{peter} attributes the scaling violations to the instantaneous
nature of the propagator considered here.

\begin{figure}
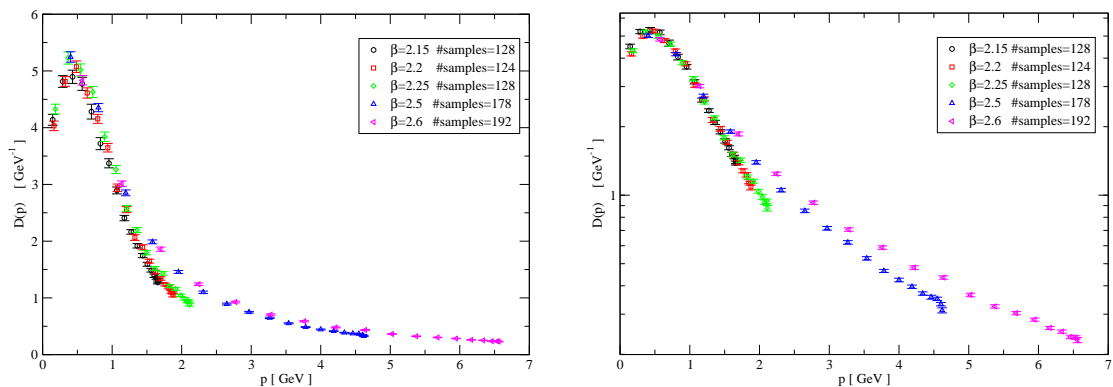

\vspace*{1mm}
\includegraphics[width=7cm]{Dp36_v7.eps}
\hspace{.5cm}
\includegraphics[width=7cm]{log_renorm_Gp36_v2.eps}
\label{fig3}
\caption{Left panel: The equal-time gluon propagator for various values
of the coupling constant. The gauge fixing includes preconditioning
and a minimum of 30 Gribov repetitions for each measurement; multiplicative
renormalisation focused on the IR data. Right panel: The same data
in a logarithmic plot.}
\end{figure}

%%%%%%%%%%%%%%%%%%%%%%%%%%%%%%%%%%%%%%%%%%%%%%%%%%%%%%%%%%%%%%%%%%%%%%%%%%%%%%%%

\section{Summary and conclusions}

In this talk, I have presented first results for the equal-time
gluon propagator measured in an improved Coulomb gauge fixing scheme.
The general observation is a significant \emph{suppression} of the
propagator in the infrared, and a \emph {loss of scaling} at very large
momenta. Although the numerics is not fully compelling, the IR data points to
$D(0) = 0$ as a likely scenario even for $D=3+1$. The failure of
multiplicative renormalisation in the UV has also been observed in
other studies treating Coulomb gauge with improved gf.~techniques;
in perturbation theory, this failure can presumable be attributed to
a loss of covariance for the equal-time propagator.

To make the present numbers more convincing, we have to go to smaller
momenta, which may involve a simulated annealing step in the gf.~pipeline.
To get a handle on the scaling issue, it would also be interesting to study
the gluon propagator at non-equal times, using a complete gauge fixing
that also destroys the residual symmetry in Coulomb gauge.
Further investigations involve the ghost propagator and the Coulomb
form factor, which are of immediate relevance for the physics of the
gauge system. These studies are currently underway and will be presented
elsewhere.

%%%%%%%%%%%%%%%%%%%%%%%%%%%%%%%%%%%%%%%%%%%%%%%%%%%%%%%%%%%%%%%%%%%%%%%%%%%%%%%%


\begin{thebibliography}{99}
\addtolength{\itemsep}{-6pt}
  \bibitem{claus_hugo} C.~Feuchter and H.~Reinhardt, Phys.~Rev.~\textbf{D70}
  (2004) 105021; \\
  H.~Reinhardt and C.~Feuchter, Phys.~Rev.~\textbf{D71} (2005) 105002.
  \bibitem{szepaniak}A.P.~Szczepaniak and E.S.~Swansen,
   Phys.~Rev.~\textbf{D65} (2002) 025015.
  \bibitem{depple} D.~Epple, H.~Reinhardt and W.~Schleifenbaum,
   Phys.~Rev.~\textbf{D75} (2007) 045011; \\
   H.~Reinhardt, W.~Schleifenbaum, D.~Epple, C.~Feuchter,
   \texttt{arXiv:0710.0316 [hep-th]}
  \bibitem{kurt} K.~Landfeld and L.~Moyaerts, Phys.~Rev.~\textbf{D70}
   (2004) 074507.
  \bibitem{cucc} A.~Cucchieri and D.~Zwanziger, Phys.~Rev.~\textbf{D65} (2002)
   014001.
  \bibitem{voigt} A.~Voigt et.~al.~, \texttt{arXiv:0709.4585 [hep-lat]}
  \bibitem{mueller_preusker} I.L.~Bogolubsky et.~al.~,
   Phys.~Rev.~\textbf{D74} (2006) 034503.
  \bibitem{zwanziger} D.~Zwanziger, Nucl.~Phys.~\textbf{B518} (1998) 237.
  \bibitem{peter} P.~Watson and H.~Reinhardt, \texttt{arXiv:0709.0140 [hep-th]}.
\end{thebibliography}
\end{document}